\documentclass[aip,10pt]{revtex4-1}
\usepackage{graphicx}
\usepackage{dcolumn}
\usepackage{bm}
\usepackage{amsmath}
\usepackage{amssymb}
\usepackage{wrapfig}
\usepackage{booktabs}
\usepackage{xcolor}
\usepackage[utf8]{inputenc}
\usepackage[T1]{fontenc}
\usepackage{mathptmx}
\usepackage{url}
\usepackage{hyperref}
\hypersetup{colorlinks=true,citecolor=blue,linkcolor=blue,urlcolor=blue}
\usepackage{lmodern}

\makeatletter
\def\@email#1#2{%
 \endgroup
 \patchcmd{\titleblock@produce}
  {\frontmatter@RRAPformat}
  {\frontmatter@RRAPformat{\produce@RRAP{*#1\href{mailto:#2}{#2}}}\frontmatter@RRAPformat}
  {}{}
}%
\makeatother

\begin{document}
\preprint{AIP/123-QED}
\title{Physics-Informed Neural Network Surrogate for Oxygen Vacancy Dynamics in epitaxial \texorpdfstring{$\mathrm{SrTiO_3}$}{SrTiO3} on Si memristors via Dynamic Spectral Optimization}
\author{R. Podorozhny}
\email{rp31@txstate.edu}
\affiliation{ 
Dept. of Computer Science, Texas State University, San Marcos, TX, USA
}%

\author{N. Theodoropoulou}
\email{ntheo@txstate.edu}
\affiliation{ 
Dept. of Physics, Texas State University, San Marcos, TX, USA
}%

\author{J. Te\v{s}i\'c}
 \email{jtesic@txstate.edu}
\affiliation{ 
Dept. of Computer Science, Texas State University, San Marcos, TX, USA
}%

\date{\today}%

\begin{abstract}
Physics-informed neural networks (PINNs) offer a promising framework for modeling semiconductor devices, yet standard architectures struggle with severe numerical stiffness and multiscale spatial discrepancies inherent to oxide heterostructures. Here, we demonstrate a cascaded PINN architecture coupled with a custom second-order Chebyshev second kind polynomial spectral optimizer (\emph{DSO V2 Hybrid}) to model ion-electronic drift-diffusion transport in \texorpdfstring{$\mathrm{Pt/SrTiO_3/Si}$}{Pt/SrTiO3/Si} memristive heterostructures across a \texorpdfstring{$20\,\text{nm}$}{20 nm} STO film on a \texorpdfstring{$380\,\mu\text{m}$}{380 micron} Si substrate. By isolating potential, carrier density, and vacancy transport into four sequentially trained sub-neural-networks, our model circumvents condition numbers exceeding \texorpdfstring{$10^{16}$}{1e16} without operator splitting. The trained surrogate reproduces experimental conductive-AFM current--voltage hysteresis (\texorpdfstring{$R^2 > 0.96$}{R2 > 0.96}) while ensuring strict Poisson consistency across continuous space. Compared to conventional finite-element solvers (e.g., COMSOL), the PINN surrogate enables differentiable inverse parameter estimation and linear time inference. 
\end{abstract}

\maketitle

\section{Introduction and Related Work}
\label{sec-intro}

Physics-informed neural networks (PINNs) solve differential equations by embedding governing physical laws directly into deep neural network loss functions \cite{raissi2019physics,karniadakis2021pinn}. Unlike purely data-driven models, PINNs enforce partial differential equations (PDEs), initial conditions, and boundary conditions by penalizing residuals at collocation points \cite{karniadakis2021pinn,lu2021deepxde}. Theoretical error analysis studies link PDE residual losses to true approximation bounds under specific assumptions \cite{deryck2022error,liu2023res,qian2023error}. In practice, standard residual minimization routinely fails in stiff or strongly non-linear physical regimes \cite{raissi2019physics,karniadakis2021pinn,wang2022bias,jagtap2020xpinn}. Existing theoretical error bounds rely primarily on simplified linear assumptions \cite{liu2023res,deryck2022error} and lack practical validation in complex, parameterized settings. Also, the PINNs are mostly validated via global metrics rather than systematic parameter probing; as a result, aggregate metrics often mask localized failures outside nominal training domains. Thus, the theoretical and practical understanding of how well PINNs approximate the true PDE solution remains incomplete, particularly when it comes to quantifying and localizing the discrepancy between the learned surrogate and high-fidelity numerical solvers \cite{qian2023error,jagtap2020xpinn}.

To address these limitations, we evaluate PINN generalization relative to reference solvers across physical parameter spaces, focusing on oxygen-vacancy concentrations. Rather than relying solely on loss values, we systematically map functional discrepancies in predicted fields and derived observables against conventional numerical console outputs. This approach provides fine-grained diagnostics of surrogate reliability, links parameter shifts directly to residual behavior, and informs better sampling strategies and network architectures. As a proof of concept, we examine the multiscale drift-diffusion transport physics in $\mathrm{SrTiO_3/Si}$ memristive heterostructures. Modeling this system requires coupling electronic drift-diffusion with mobile oxygen-vacancy dynamics. Dynamic ionic-electronic coupling, localized space-charge accumulation, and exponential field-dependent drift terms introduce severe numerical stiffness near the interfaces.

\textbf{Related Work:} Physics-informed neural networks embed PDE residual terms into loss functions for forward and inverse boundary value problems \cite{raissi2019physics,karniadakis2021pinn,lu2021deepxde,jagtap2020xpinn}. Modern theoretical frameworks establish rigorous relationships between PDE loss residuals and true solution error bounds \cite{deryck2022error,qian2023error,liu2023res}. Mitigating optimization failures in highly non-linear or multiscale PDE regimes remains an active area of investigation \cite{wang2022bias,HoddPINN2023}. The present work extends these methodologies by providing end-to-end differentiable surrogates that outperform classical finite-element numerical solvers across extreme parameter scales.

\section{Physical System}
\label{sec-physics}

We model electronic transport and ionic migration in a metal--oxide--semiconductor memristive heterostructure consisting of a $20\,\text{nm}$ strontium titanate ($\mathrm{SrTiO_3}$, STO) thin film deposited on a $380\,\mu\text{m}$ silicon (Si) substrate with a platinum (Pt) top Schottky contact ($\mathrm{Pt/SrTiO_3/Si}$) \cite{kollias2025oxygen}. Reversible resistive switching in this system is driven by electric-field-induced electromigration (drift and diffusion) of mobile donor oxygen vacancies ($\mathrm{V_O^{2+}}$ or $V_{\mathrm{O}}^{\bullet\bullet}$) \cite{kollias2025oxygen}. Conductive-atomic force microscopy (C-AFM) experiments confirm that applied bias voltages induce redistribution of these oxygen vacancies across the $20\,\text{nm}$ STO film \cite{kollias2025oxygen}. This dynamic spatial redistribution modulates the interfacial space-charge profile, band bending, and effective Schottky barrier height at the $\mathrm{Pt/STO}$ contact. Accurately modeling these transport dynamics requires directly resolving the non-linear coupling between ionic space-charge buildup and exponential electron injection.

To describe the continuous physics of coupled ionic-electronic transport, the spatial domain is mapped to a normalized coordinate $x \in [0, 1]$, where $x = 0$ corresponds to the $\mathrm{Pt/STO}$ Schottky interface, $x_i = 0.5$ defines the internal $\mathrm{STO/Si}$ heterojunction, and $x = 1.0$ represents the back substrate contact. Electrostatic potentials are non-dimensionalized by the thermal voltage $U_t = k_B T / q \approx 25.9\,\text{mV}$ at $T = 300\,\text{K}$, and local carrier concentrations are normalized to the nominal donor density of STO, $N_{D,\mathrm{STO}} = 10^{18}\,\text{cm}^{-3}$. Two coupled partial differential equations govern the self-consistent physical model:

 \noindent \textbf{1. Non-linear Poisson Equation for Electrostatic Potential:} The normalized electrostatic potential $\varphi(x)$ satisfies: \begin{equation}\label{eq-poisson} \lambda^2\,\frac{\partial^2 \varphi}{\partial x^2} \;=\; Z_v\,C_v \;-\; n \;+\; p\,\end{equation} The $\lambda^2 = (\lambda_D / L)^2 \approx 1.07\times 10^{-1}$ is the dimensionless Debye parameter ($\lambda_D \approx 6.55\,\text{nm}$ in STO, $L = 20\,\text{nm}$ film thickness), $Z_v = 2$ is the vacancy charge state, $C_v(x,t)$ is the normalized vacancy concentration, and $n(x,t), p(x,t)$ are normalized electron and hole concentrations.

\noindent \textbf{2. Oxygen-Vacancy Drift--Diffusion Continuity Equation:} The temporal evolution of the mobile donor vacancy concentration $C_v(x,t)$ is governed by:  \begin{equation}\label{eq-vacancy}  \frac{\partial C_v}{\partial t} \;=\; D_v\,\frac{\partial^2 C_v}{\partial x^2} \;-\; Z_v\,\mu_v\,\frac{\partial}{\partial x}\left(C_v\,\frac{\partial\varphi}{\partial x}\right),  \end{equation} The $D_v = 0.01$ and $\mu_v = 0.05$ represent normalized vacancy diffusivity and mobility, respectively. The drift term couples ionic transport directly to the local electrostatic electric field $E = -\partial\varphi/\partial x$.

\noindent \textbf{Boundary and Initial Conditions:}
The coupled PDE system is subjected to the following physical boundary and initial constraints. The $\Phi_B = 1.3\,\text{eV}$ is the zero-bias Schottky barrier height.
\begin{align}
\varphi(x=0) &= \Phi_{B,\mathrm{norm}} + \frac{V(t)}{U_t} & &\text{(Schottky contact under applied bias $V$)}\,,\label{eq-bc0}\\
\varphi(x=x_i) &= 0 & &\text{(STO/Si heterojunction reference, $x_i = 0.5$)}\,,\label{eq-bcintf}\\
C_v(x, t=0) &= 1 & &\text{(Equilibrium initial vacancy distribution)}\,,\label{eq-ic}
\end{align}

\section{COMSOL Modeling Limitations}
\label{sec-Limits}

Classical finite-element numerical solvers (such as COMSOL Multiphysics using its Semiconductor Module) represent the state of the art for modeling 1D transport, energy band diagrams, and current--voltage ($I\text{--}V$) hysteresis in the $\mathrm{Pt/STO/Si}$ system \cite{kollias2025oxygen,kollias2018thesis}. Previous numerical studies constructed 1D models incorporating parabolic energy bands, Maxwell-Boltzmann statistics, and active oxygen vacancy distributions within the $20\,\text{nm}$ STO film on an n-type silicon substrate \cite{kollias2025oxygen,kollias2018thesis}. By shifting the position of the vacancy layer between the interface (low-resistance state/trace sweep) and the film surface (high-resistance state / retrace sweep), these models successfully reproduced qualitative $I\text{--}V$ hysteresis, zero-current voltage shifts, and rectifying diode behavior, establishing that resistive switching in these heterojunctions is driven by vacancy migration and barrier modulation \cite{kollias2025oxygen,kollias2018thesis}.

However, conventional grid-based finite-element discretizations encounter severe physical and computational bottlenecks when modeling coupled ionic-electronic transport in this heterostructure:

\begin{enumerate}
    \item \textbf{Multiscale Spatial Disparity and Domain Truncation:} A five-order-of-magnitude length-scale disparity exists between the Debye screening length ($\lambda_D \approx 6.55\,\text{nm}$ in STO), the STO film thickness ($L = 20\,\text{nm}$), and the physical bulk Si substrate ($380\,\mu\text{m}$) \cite{kollias2025oxygen,kollias2018thesis}. Finite-element meshes require intense spatial refinement near interfaces to resolve narrow space-charge layers. To maintain tractable element counts and prevent prohibitive runtimes, standard models artificially truncate the Si substrate thickness to $10\,\mu\text{m}$ and simplify 2D/3D tip-sample contact geometries to 1D approximations \cite{kollias2025oxygen,kollias2018thesis}.

    \item \textbf{Static Discrete Vacancy Approximations vs.\ Dynamic Continuity:} Rather than solving the fully time-dependent drift-diffusion continuity equation for mobile oxygen vacancies during dynamic voltage sweeps, COMSOL workflows approximate ionic redistribution using static, discrete $5\,\text{nm}$ charged blocks ($5\times 10^5\,\text{C/cm}^3$) fixed near either the $\mathrm{Pt/STO}$ interface or the film surface \cite{kollias2025oxygen,kollias2018thesis}. This prevents continuous, self-consistent modeling of electric-field-driven electromigration kinetics.

    \item \textbf{Numerical Stiffness and Mesh Non-Convergence:} Non-linear coupling between electrostatic potential and carrier distributions introduces mathematical stiffness. The Debye parameter ($\lambda^2 = (\lambda_D/L)^2 \approx 1.07\times 10^{-1}$), roughly an order of magnitude smaller than the normalized charge-density term it must balance against, yields sharp field gradients across thin depletion regions. Standard Newton--Raphson mesh solvers suffer from ill-conditioning, frequent non-convergence, or impractically long iteration runtimes, rendering continuous dynamic voltage sweeps and multi-parameter explorations (e.g., varying STO donor doping $N_{D,\mathrm{STO}} = 10^{18}\,\text{cm}^{-3}$) computationally prohibitive \cite{kollias2025oxygen,kollias2018thesis}.

    \item \textbf{Simplified Statistics and Idealized Physics:} Standard finite-element solvers adopt idealized parabolic energy bands, complete dopant ionization, and Maxwell-Boltzmann carrier statistics (instead of full Fermi-Dirac statistics) to reduce non-linear numerical complexity and maintain solver stability \cite{kollias2025oxygen}.

    \item \textbf{Non-Differentiable Forward-Only Parameter Fitting:} Finite-element tools operate as forward-only solvers. Outer-loop parameter fitting (e.g., extracting series resistance, ideality factors, or barrier modulation coefficients) requires thousands of independent PDE evaluations, making systematic inverse optimization computationally intractable \cite{kollias2018thesis}.
\end{enumerate}

To overcome the mesh stiffness, substrate truncation, and convergence failures of classical grid-based solvers, Section~\ref{sec-pinn} presents a physics-informed machine learning framework incorporating log-space loss transformations for the resulting scale mismatch in Debye screening ($\lambda^2 \approx 0.107$), sequential multi-stage sub-network architectures, and continuous PDE residual minimization.

\section{Cascaded Physics Informed Neural Sub-Networks Design}
\label{sec-pinn}

\begin{figure}[!ht]
\centering
\includegraphics[width=0.85\textwidth]{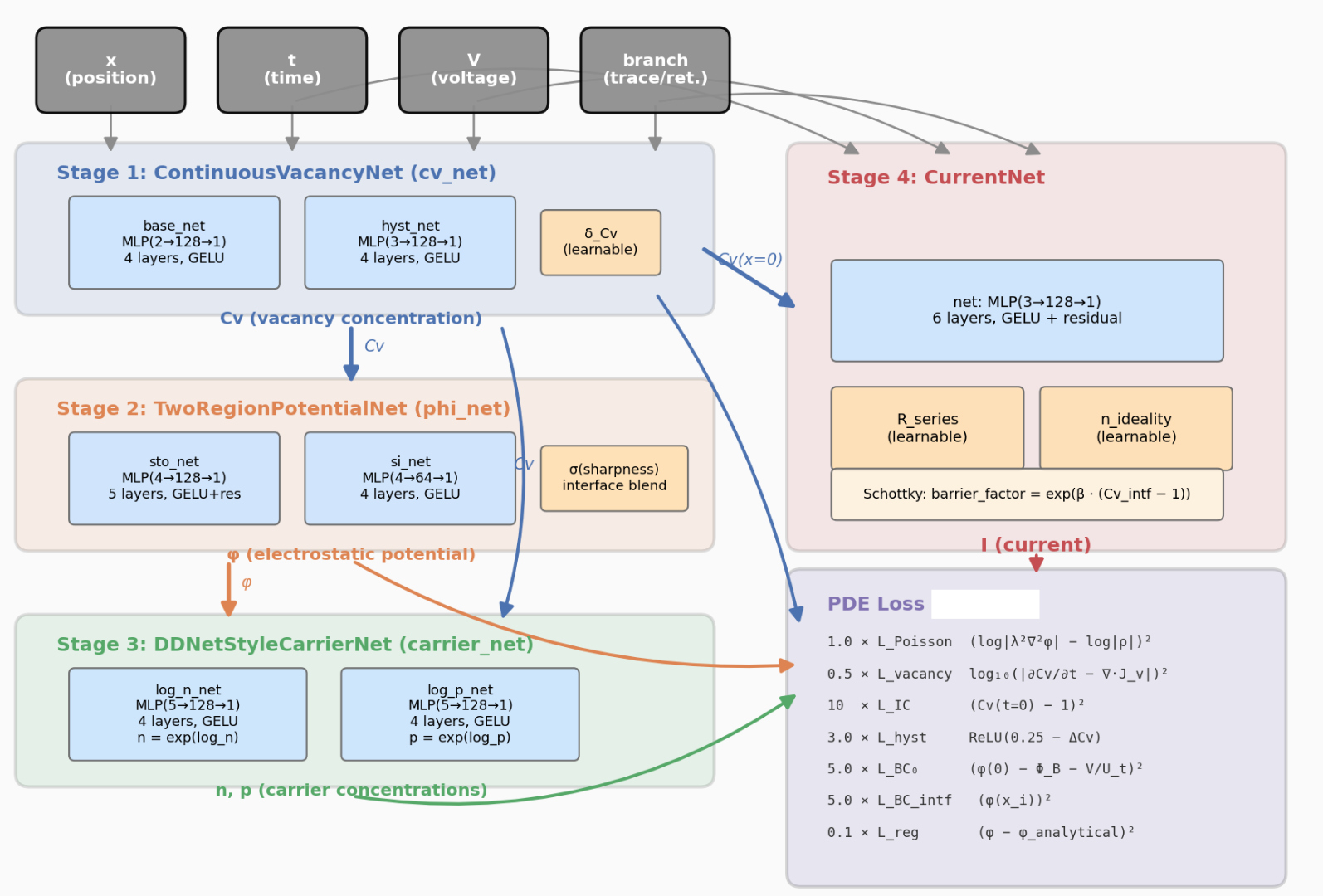}
\caption{Cascaded architecture of the $\mathrm{Pt/STO/Si}$ memristor PINN. Four sub-neural-networks are evaluated in sequence: Stage 1 (\emph{$cv_{net}$}) models $C_v$; Stage 2 (\emph{$\varphi_{net}$}) maps electrostatic potential $\varphi$; Stage 3 (\emph{$carrier_{net}$}) computes carrier concentrations $(n,p)$; Stage 4 (\emph{$current_{net}$}) calculates total terminal current $I$. All PDE-coupled fields feed into the multi-objective loss evaluation.}
\label{fig-architecture}
\end{figure}

We propose a novel PINN architecture that embeds physical domain knowledge directly into the network's inductive biases, log-space loss transformations, and modular sub-neural-networks \cite{raissi2019physics,karniadakis2021pinn}. The proposed PINN architecture is illustrated in Figure~\ref{fig-architecture}. To decouple complex multi-physics interactions and eliminate the severe numerical stiffness ($\kappa(\mathbf{J}) \sim 10^{16}$, $\kappa$ - Hessian condition number) caused by coupled ionic-electronic dynamics, we introduce a directional causality cascade composed of four sequentially evaluated sub-neural-networks. This architecture enforces physical dependencies while isolating numerical instabilities, enabling independent sub-network pretraining, domain-aligned network topologies, selective parameter freezing, and strict gradient isolation. As illustrated in Figure~\ref{fig-architecture}, each sub-network embeds tailored physical inductive biases along the integrated forward pass execution pipeline:
\begin{equation}
C_v = cv_{net}(x, t, V, b) \longrightarrow 
\varphi = \varphi_{net}(x, t, V, C_v) \longrightarrow 
(n, p) = carrier_{net}(x, t, V, \varphi, C_v) \longrightarrow 
I = current_{net}(t, V, b, C_v|_{x=0}).
\end{equation}

\paragraph{Stage 1: Vacancy Concentration Network (\emph{$cv_{net}$})} Physical spatial decay envelopes enforce directional vacancy accumulation during bias sweeps:
\begin{equation}
C_v^{\mathrm{base}}(x) = 1 + f_{\mathrm{base}}(x, t)\cdot
\begin{cases}
\exp(-5x) & \text{trace sweep } (b = 0),\\
\exp\!\bigl(-5(1-x)\bigr) & \text{retrace sweep } (b = 1).
\end{cases}
\end{equation}
The class \emph{ContinuousVacancyNet} maps spatial coordinate $x$, time $t$, voltage $V$, and sweep direction $b \in \{0,1\}$ (trace/retrace) to local vacancy density $C_v$. Dynamic hysteresis injected by\emph{hyst\_net} is scaled by $|\delta_{C_v}|\,t$, enforcing monotonic memory accumulation over time. The scalar output is clamped to $C_v \in [0.3, 5.0]$ for numerical stability. The components of the \emph{$cv_{net}$} are: 
\begin{table}[!ht]
\centering
\setlength\tabcolsep{1pt}
\begin{tabular}{lll}
\textbf{Component} & \textbf{Topology} & \textbf{Physical Role / Notes} \\
\emph{base\_net} & MLP: $2 \to 128 \to 128 \to 128 \to 1$ & 4 layers, GELU, residual connections \\
\emph{hyst\_net} & MLP: $3 \to 128 \to 128 \to 128 \to 1$ & 4 layers, GELU, models dynamic trace-retrace deviation \\
$\delta_{C_v}$ & Scalar (init. $0.5$) & Learnable dynamic hysteresis amplitude \\
\end{tabular}
\end{table}

\paragraph{Stage 2: Electrostatic Potential Network (\emph{$\varphi_{net}$})}
The class \emph{TwoRegionPotentialNet} resolves heterojunction potential distributions across the material boundary using region-specific sub-neural-networks blended via a learnable sigmoid function. The blended electrostatic potential field is constructed as: $
\varphi(x) = \sigma\!\bigl(s\,(x - x_i)\bigr)\,\varphi_{\mathrm{STO}}(x) + \bigl[1 - \sigma\!\bigl(s\,(x - x_i)\bigr)\bigr]\,\varphi_{\mathrm{Si}}(x)\,$
where $x_i = 0.5$. The STO potential is anchored directly to the Schottky contact boundary condition with a vacancy-modulated correction $-0.5\,(C_v - 1)$, capturing Schottky barrier lowering under vacancy accumulation. The components of the \emph{$\varphi_{net}$} are: \begin{table}[!ht]
\begin{tabular}{lll}
\centering
\setlength\tabcolsep{1pt}
\textbf{Component} & \textbf{Topology} & \textbf{Physical Role / Notes} \\
\midrule
\emph{sto\_net} & MLP: $4 \to 128^{\times 4} \to 1$ & 5 layers, GELU, residual, models STO layer physics \\
\emph{si\_net} & MLP: $4 \to 64^{\times 3} \to 1$ & 4 layers, GELU, models Si space-charge layer \\
$s$ (sharpness) & Scalar (init. $200$) & Learnable sigmoid boundary blending width \\
\bottomrule
\end{tabular}
\end{table}

\paragraph{Stage 3: Carrier Concentration Network (\emph{$carrier_{net}$})}
The class \emph{DDNetStyleCarrierNet} computes electron ($n$) and hole ($p$) concentrations in log-space relative to their analytical drift-diffusion equilibrium profiles. The baseline equilibrium contribution is computed analytically: $\log n_{\mathrm{eq}}^{\mathrm{STO}} = \log C_v - \frac{\varphi - \varphi_0}{U_t}\,$
and the network predicts a learned non-equilibrium departure $\Delta\log n$, such that $n = \exp\bigl(\log n_{\mathrm{eq}} + s_n\,\Delta\log n\bigr)$, where $s_n$ is a learnable scaling parameter (initialized to $3.0$). Values are clamped to $[-30, 30]$ before exponentiation to prevent overflow. The components of the \emph{$carrier_{net}$} are: \begin{table}[!ht]
\begin{tabular}{lll}
\centering
\setlength\tabcolsep{1pt}
\textbf{Component} & \textbf{Topology} & \textbf{Physical Role / Notes} \\
\midrule
\emph{log\_n\_net} & MLP: $5 \to 128^{\times 3} \to 1$ & 4 layers, GELU, computes non-equilibrium electron deviation \\
\emph{log\_p\_net} & MLP: $5 \to 128^{\times 3} \to 1$ & 4 layers, GELU, computes non-equilibrium hole deviation \\
\bottomrule
\end{tabular}
\end{table}

\paragraph{Stage 4: Current Network (\emph{$current_{net}$})}
The class \emph{CurrentNet} maps interfacial state variables $(t, V, b, C_v|_{x=0})$ directly to total device current $I$, explicitly accounting for Schottky barrier modulation and parasitic series resistance. Interfacial vacancy accumulation modulates the effective Schottky barrier through an exponential enhancement factor: $f_{\mathrm{barrier}} = \exp\!\bigl(|\beta|\,(C_v|_{x=0} - 1)\bigr)$. The predicted base current is scaled by $f_{\mathrm{barrier}}$ and a forward-bias factor $1 + 2\,\mathrm{ReLU}(V)$, followed by an inline series-resistance correction $V_{\mathrm{eff}} = V - I R_{\mathrm{series}}$. The components of the \emph{$current_{net}$} are: \begin{table}[!ht]
\begin{tabular}{lll}
\centering
\setlength\tabcolsep{1pt}
\textbf{Component} & \textbf{Topology} & \textbf{Physical Role / Notes} \\
\midrule
\emph{net} & MLP: $3 \to 128^{\times 5} \to 1$ & 6 layers, GELU, residual architecture \\
$R_{\mathrm{series}}$ & Scalar (init. $100\,\Omega$) & Learnable parasitic series resistance \\
$n_{\mathrm{trace/ret}}$ & Scalars ($1.5$, $1.8$) & Learnable diode ideality factors per sweep branch \\
$\beta$ & Scalar (init. $3.0$) & Learnable Schottky barrier modulation coefficient \\
\bottomrule
\end{tabular}
\end{table}

\section{Methodology}
\label{sec-method}

\begin{figure*}[!ht]
\includegraphics[width=0.7
\linewidth]{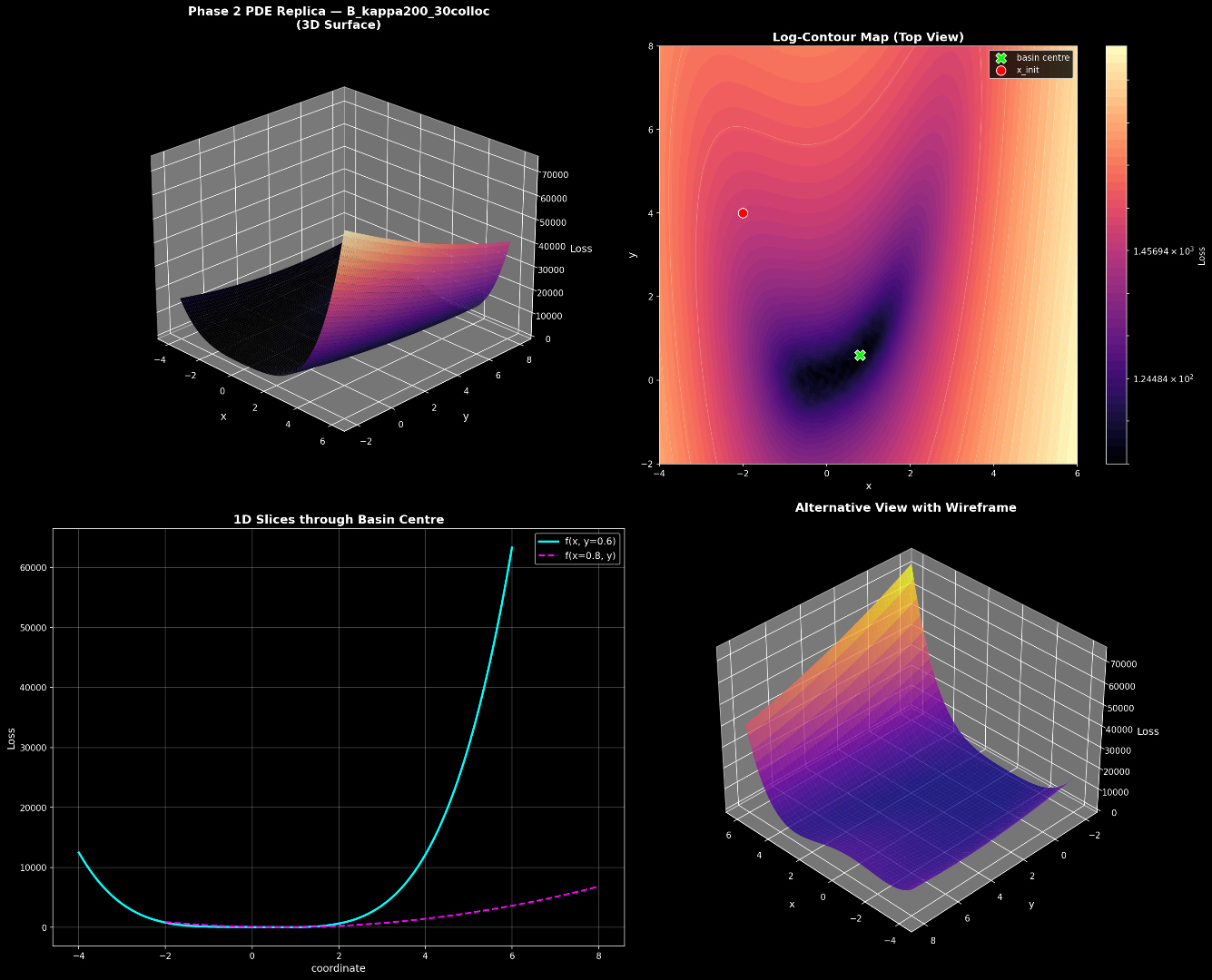}
\caption{Loss landscape for the \emph{$carrier_{net}$} loss}
\label{fig-PDEloss}
\end{figure*}

Physics-informed neural networks (PINNs) generate stiff, highly non-convex loss landscapes dominated by extreme condition numbers ($\kappa \sim 10^{16}$), dense multi-saddle geometries, and severe spectral anisotropy \cite{raissi2019physics}. If the loss landscape has cross-coupled ill-conditioning and/or rich saddles, Adam's diagonal preconditioning struggles on those landscapes \cite{podorozh2026adamstall}. 

State-of-the-art optimizers in deep learning remain overwhelmingly first-order frameworks, encompassing both momentum-based optimizers such as Adam \cite{kingma2015adam} and pseudo-second-order preconditioned optimizers \cite{gupta2018shampoo,vyas2024soap}. It is mainly so due to their great scalability and robustness. Under physics-informed neural network (PINN) conditions, first-order methods may fail because they struggle on cross-coupled ill-conditioned or saddle rich loss landscapes, leading to severe numerical oscillations, gradient pathologies, and premature training stagnation \cite{podorozh2026adamstall}. Meanwhile, pseudo-second-order optimizers \cite{gupta2018shampoo,vyas2024soap} construct positive-semidefinite, Kronecker-factored preconditioners derived from gradient outer products to approximate tensor curvature without the $O(n^3)$ cost of explicit Hessian inversions. Since the preconditioners are positive-semidefinite \cite{gomes2025adafisher,abreu2026gn}, they remain incapable of capturing negative curvature to escape multi-saddle points \cite{dauphin2014saddle} or resolving extreme, non-separable ill-conditioning without numerical instability.

Second-order optimizers may outperform diagonal adaptive methods such as Adam~\cite{kingma2015adam} when navigating surfaces characterized by cross-coupled ill-conditioning \cite{nesterov2006cubic,martens2010icml,kiyani2025which,podorozh2026adamstall}. Some second-order optimization frameworks offer theoretical minimization guarantees under cubic regularization bounds on the loss function \cite{nesterov2006cubic}. However, extracting exact curvature requires direct matrix solves with a prohibitive $O(n^3)$ computational cost, rendering theoretical Newton updates intractable on non-convex landscapes beyond 15k--20k parameters. To bypass the bottleneck, research leverages blockwise decomposition, Krylov subspace methods, or Chebyshev polynomial recurrences \cite{podorozh2026blockarcphi1} alongside fast curvature estimation techniques, including Hutchinson trace estimation, Kronecker factorization, and exponential moving average (EMA)-smoothed Hessian-vector products (HVPs) \cite{pearlmutter1994hessian}. These measures may reduce computational time to $O(Ln)$ (where $L$ is a curvature dependent degree used in some HVP based methods) while maintaining sufficient second-order fidelity, enabling cubic-regularized algorithms to navigate complex loss geometries at significantly larger parameter scales \cite{podorozh2026blockarcphi1}. 

Figure~\ref{fig-PDEloss} illustrates that the drift-diffusion Poisson system produces an exceptionally ill-conditioned loss landscape where severe spectral anisotropy and stiff gradient coupling exacerbate these computational challenges.
One of the ways to mitigate these difficulties in this work was the use of some variants of a second-order optimizer, DSO~\cite{podorozhny2026}.

\subsection{Hybrid Dynamic Spectral Optimizer}
A Dynamic Spectral Optimizer (\emph{DSO}) was evaluated on some subnets of the PINN. It utilizes a polynomial recurrence, specifically Chebyshev polynomial of the second kind, to rescale the eigenvalue spectrum of an ill-conditioned Hessian for uniform contraction rate along all eigenvalue directions. Since it is implemented as a recurrence, it approximates a Hessian inversion via matrix by vector multiplications, avoiding a linear solve \cite{podorozhny2026dso}. A number of variants of this optimizer were built that used different update rules based on the size of a tensor. One of the variants, \emph{DSO V2 Hybrid} \cite{podorozhny2026,podorozhny2026dso}, used two update rules based on the tensor size. DSO V2 Hybrid dynamically adapts the degree of Chebyshev polynomial based on Hessian spectral properties.

\paragraph{Adaptive Relaxation Selection.} For instance, the loss landscape of \emph{$current_{net}$} with a large tensor is characterized by dynamic exponential transport terms with steep gradients; consequently, an exponential relaxation function is applied that could drive the loss to very low levels, as shown in Table~\ref{tab-pretrainingResults}.

\paragraph{Carrier Concentration Log-Space Formulation} 

Because small perturbations in electrostatic potential $\varphi$ trigger exponential variations in carrier concentrations ($n$ and $p$), the resulting loss landscape exhibits extreme Hessian anisotropy and dense cross-coupled curvature, yielding condition numbers on the order of $\kappa \sim 10^{16}$. To resolve extreme physical dynamic ranges, carrier concentrations $n$ and $p$ are represented in exponential log-space formulations as  $n = \exp\!\left(\log n_{\mathrm{eq}} + s_n\,\Delta\log n\right)$, and $p = \exp\!\left(\log p_{\mathrm{eq}} + s_p\,\Delta\log p\right)$. The equilibrium electron concentration is coupled to the electrostatic potential as $\log n_{\mathrm{eq}}^{\mathrm{STO}} = \log C_v - \frac{\varphi - \varphi_0}{U_t}, \qquad U_t = 25.9\,\mathrm{mV}$. 

\paragraph{Dynamic Block Partitioning \& Curvature Smoothing.} Parameter blocks, tensor dimensions, and exponential moving-average (EMA) smoothing parameters for curvature tracking are dynamically updated to preserve second-order sensitivity without suffering from gradient noise or ill-conditioning.

\paragraph{Sub-network-Specific Optimizer Performance.} Because each sub-network in our multi-stage PINN architecture exhibits a distinct loss landscape, DSO V2 Hybrid showed good performance on highly ill-conditioned loss landscapes.

\subsection{Composite PDE Training Loss} 

The continuous physics of coupled ionic-electronic transport is governed by two normalized PDEs operating over $x \in [0, 1]$, where $x=0$ is the $\mathrm{Pt/STO}$ interface and $x_i = 0.5$ is the $\mathrm{STO/Si}$ junction explained in Section~\ref{sec-physics}. The normalized electrostatic potential $\varphi(x)$ satisfies Eq~\ref{eq-poisson}. Because $\lambda^2 \approx 0.107$ is an order of magnitude smaller than the normalized reaction term $\rho$, standard linear residuals $\mathcal{L} = \langle (\lambda^2 \partial_{xx}\varphi - \rho)^2 \rangle$ suffer from imbalanced gradient contributions and numerical ill-conditioning. We overcome this by defining the residual loss in log-space as in Eq.~\ref{eq-poissonLoss}. The $\varepsilon_{\mathrm{num}} = 10^{-10}$ ensures numerical stability.
    \begin{equation}\label{eq-poissonLoss}
    \mathcal{L}_{\mathrm{Poisson}} \;=\; \Big\langle \Bigl( \log_{10}\!\bigl(|\lambda^2\,\partial_{xx}\varphi| + \varepsilon_{\mathrm{num}}\bigr) - \log_{10}\!\bigl(|\rho| + \varepsilon_{\mathrm{num}}\bigr) \Bigr)^2 \Big\rangle_{\mathbf{x}}\,,\qquad \rho = Z_v C_v - n + p\,,
    \end{equation}
    
    The temporal evolution of mobile donor vacancies $C_v(x,t)$ is governed by Eq.\ref{eq-vacancy} explained in Section~\ref{sec-physics}. The drift residual is also represented as a loss function in log-space as 
    \begin{equation}\label{eq-vacancyLoss}
    \mathcal{L}_{\mathrm{vac}} \;=\; \Big\langle \Bigl( \log_{10}\!\bigl(|r_{\mathrm{vac}}| + \varepsilon_{\mathrm{num}}\bigr) \Bigr)^2 \Big\rangle_{\mathbf{x}}\,,\qquad r_{\mathrm{vac}} = \partial_t C_v - D_v\,\partial_{xx}C_v + Z_v\mu_v\bigl(\partial_x C_v\,\partial_x\varphi + C_v\,\partial_{xx}\varphi\bigr).
    \end{equation}
    
We propose that the composite PDE training loss balances mathematical residual satisfaction, boundary conditions, physical regularizations, and hysteresis constraints:
\begin{equation}\label{eq-totalLoss}
\mathcal{L}_{\mathrm{total}} \;=\; \underbrace{1.0\,\mathcal{L}_{\mathrm{Poisson}} + 0.5\,\mathcal{L}_{\mathrm{vac}}}_{\text{PDE Physical Residuals}} \;+\; \underbrace{10\,\mathcal{L}_{\mathrm{IC}} + 3\,\mathcal{L}_{\mathrm{hyst}}}_{\text{Initial \& Hysteresis}} \;+\; \underbrace{5\,\mathcal{L}_{\mathrm{BC}_0} + 5\,\mathcal{L}_{\mathrm{BC}_i}}_{\text{Boundary Constraints}} \;+\; \underbrace{0.1\,\mathcal{L}_{\mathrm{reg}}}_{\text{Analytical Regularization}}
\end{equation}
 The individual sub-losses are defined as:
\begin{align}
\mathcal{L}_{\mathrm{IC}} &= \bigl\langle \bigl(C_v(x, 0) - 1\bigr)^2 \bigr\rangle\,,\\
\mathcal{L}_{\mathrm{hyst}} &= \bigl\langle \mathrm{GELU}(0.25 - \Delta C_v)\bigr\rangle\,,\qquad \Delta C_v = C_v^{\mathrm{trace}}(0,t) - C_v^{\mathrm{retrace}}(0,t)\,,\\
\mathcal{L}_{\mathrm{BC}_0} &= \Big\langle \bigl(\varphi(0) - \Phi_{B,\mathrm{norm}} - V/U_t\bigr)^2 \Big\rangle\,,\\
\mathcal{L}_{\mathrm{BC}_i} &= \langle \varphi(x_i)^2 \rangle\,,\\
\mathcal{L}_{\mathrm{reg}} &= 0.1\,\Big\langle \bigl(\varphi - \varphi_{\mathrm{analytical}}\bigr)^2 \Big\rangle\,.
\end{align}

\subsection{Network Formulation, Activation, and Training}
\label{ssec-pinn-design}

\textbf{The Training Pipeline} navigates non-convex loss landscapes and prevents unphysical local minima as follows:
\paragraph{Analytical Pretraining (7,000 epochs/sub-network).} sub-neural-networks are independently initialized against analytical targets using a novel dynamic spectral optimizer with Chebyshev second kind acceleration explained in Section~\ref{sec-method}.
\paragraph{PDE Residual Minimization (10,000 epochs).} The PDE residual loss $\mathcal{L}_{\mathrm{total}}$ is minimized over 500 Sobol collocation points. Structural stability is enforced by freezing \emph{$\varphi_{net}$} and hysteresis parameters.
\paragraph{Differentiable Experimental Fitting.} All network weights and physical parameters ($R_{\mathrm{series}}$, $n_{\mathrm{trace/ret}}$, $\beta$, $\delta_{C_v}$) are unfrozen and jointly optimized against experimental $I\text{--}V$ curves using PCGrad to resolve gradient interference between trace and retrace branches \cite{kollias2018thesis}.

\paragraph{Activation Function Selection Strategy}
Activation function selection directly dictates gradient propagation and residual accuracy: (1) \emph{ReLU ($\mathrm{ReLU}(x) = \max(0, x)$):} Unsuitable because second derivatives vanish identically ($\partial_{xx}\mathrm{ReLU} = 0$), rendering Poisson residual evaluation invalid; (2) \emph{SIREN ($\sin(\omega_0 x)$):} High-frequency oscillatory bias caused $\sim 29\times$ higher loss than GELU due to spectral mismatches with monotonic electrostatic potential profiles; and (3) \emph{GELU (Selected):} $\mathrm{GELU}(x) = x \Phi(x)$ provides $C^\infty$-smoothness, non-periodic behavior, superior Hessian conditioning, and robust non-zero second-order gradients.

\paragraph{Collocation and Automatic Differentiation} is achieved by evaluating second-order spatial derivatives ($\partial_{xx}\varphi$ and $\partial_{xx}C_v$) across $500$ Sobol collocation points. Also, we use vector-mapped automatic differentiation (rather than manual finite-difference methods) to optimize computational performance while preserving exact gradient evaluation. 

\paragraph{Physical and Learnable Parameters} are categorized into fixed material properties (Table~\ref{tab-physicalConstants}) and trainable physical parameters (Table~\ref{tab-learnableParameters}): 

\begin{table}[!ht]
\centering
\caption{Fixed Material Properties and System Constants}
\label{tab-physicalConstants}
\begin{tabular}{llll}
\hline
\textbf{Property / Parameter} & \textbf{Symbol} & \textbf{Value} & \textbf{Physical Role / Source} \\
\hline
STO Permittivity & $\varepsilon_{\mathrm{STO}}$ & 30 & Electrostatic screening \\
Si Permittivity & $\varepsilon_{\mathrm{Si}}$ & 11.7 & Substrate screening \\
STO Bandgap & $E_{g,\mathrm{STO}}$ & 3.2\,eV & Defines carrier statistics \\
Si Bandgap & $E_{g,\mathrm{Si}}$ & 1.12\,eV & Substrate bandgap \\
STO Electron Affinity & $\chi_{\mathrm{STO}}$ & 4.4\,eV & Band alignment \\
Si Electron Affinity & $\chi_{\mathrm{Si}}$ & 4.05\,eV & Interface alignment \\
Conduction Band Offset & $\Delta E_c$ & 0.35\,eV & HAXPES measurement \\
Schottky Barrier Height & $\Phi_B$ & 1.3\,eV & $\Phi_{\mathrm{Pt}} - \chi_{\mathrm{STO}}$ \\
STO Donor Density & $N_{D,\mathrm{STO}}$ & $10^{18}$\,cm$^{-3}$ & Free carrier concentration \\
Si Donor Density & $N_{D,\mathrm{Si}}$ & $5\times 10^{15}$\,cm$^{-3}$ & Background doping \\
STO Film Thickness & $L_{\mathrm{STO}}$ & 20\,nm & Device active layer \\
Vacancy Charge State & $Z_v$ & 2 & Ionization state \\
Vacancy Diffusivity (norm.) & $D_v$ & 0.01 & Ionic diffusion rate \\
Vacancy Mobility (norm.) & $\mu_v$ & 0.05 & Field-driven drift rate \\
Debye Length (STO) & $\lambda_D$ & 6.55\,nm & $\sqrt{\varepsilon_0 \varepsilon_r k T / q^2 N_D}$ \\
Debye Scaling Factor & $\lambda^2$ & 0.107 & $(\lambda_D / L)^2$ \\
Thermal Voltage & $U_t$ & 25.9\,mV & $k T / q$ at $300\,\text{K}$ \\
\hline
\end{tabular}
\end{table}

\begin{table}[!ht]
\centering
\caption{Learnable Parameters Initialized for Optimization}
\label{tab-learnableParameters}
\begin{tabular}{llll}
\hline
\textbf{Sub-network} & \textbf{Parameter} & \textbf{Initial Value} & \textbf{Physical Role} \\
\hline
\emph{$cv_{net}$} & $\delta_{C_v}$ & 0.5 & Hysteresis amplitude \\
\emph{$\varphi_{net}$} & \textnormal{interface\_sharpness} & 200 & Sigmoid blending width \\
\emph{$carrier_{net}$} & \text{log(n) scale} & 3.0 & Log-space carrier scale \\
\emph{$carrier_{net}$} & \text{log(p) scale} & 3.0 & Log-space carrier scale \\
\emph{$current_{net}$} & $R_{\mathrm{series}}$ & $100\,\Omega$ & Series resistance \\
\emph{$current_{net}$} & $n_{\mathrm{trace}}$ & 1.5 & Trace ideality factor \\
\emph{$current_{net}$} & $n_{\mathrm{retrace}}$ & 1.8 & Retrace ideality factor \\
\emph{$current_{net}$} & $\beta$ & 3.0 & Vacancy-barrier coupling \\
\hline
\end{tabular}
\end{table}

\section{Evaluation, Validation Metrics, and Performance Analysis}
\label{sec-pinn-evaluation}

\subsection{Pretraining and Loss Landscape Analysis}
Systematic empirical mapping of the loss landscapes across the PINN sub-neural-networks in Table~\ref{tab-pretrainingResults} reveals extreme numerical ill-conditioning driven by carrier exponentiation $n = n_i \exp(\varphi/U_t)$, yielding Hessian condition number up to $\sim 10^{16}$. Search trajectory analysis and Hessian based diagnostics indicate that the sub-networks exhibit substantially different condition number trajectories, negative spectral mass, and degrees of off-diagonal curvature coupling \cite{podorozh2026adamstall}. Complementary empirical Neural Tangent Kernel (NTK) diagnostics characterize differences in the positive-semidefinite Gauss--Newton (GN) kernel spectrum and the associated mode-wise first-order training dynamics. The NTK diagnostics further reveal differences in kernel evolution and mode-wise convergence. These findings validate the necessity of applying specialized hybrid optimizers across distinct sub-network loss landscapes to maximize PINN solution fidelity. The features mentioned above are future directions of the optimizer development. In this project, early variants of the DSO optimizers were evaluated on some sub-networks. In the cascaded PINN architecture, pretraining quality on \emph{$\varphi_{net}$} establishes the accuracy floor for Poisson residuals: achieving an MSE of $4.68\times10^{-3}$ is a $6.6\times$ improvement over Adam ($3.09\times10^{-2}$) and a $23\times$ improvement over the DSO-EMA baseline ($1.10\times10^{-1}$) --- a more modest gain than on the other sub-networks, where improvements reach up to $545\times$ (\emph{$current_{net}$}), as shown in Table~\ref{tab-pretrainingResults}. One of the early optimizer variants, DSO V2 Hybrid, which switched an update rule based on the tensor size, showed strong results. This is consistent with the finding of \cite{podorozh2026adamstall} that Adam's diagonal preconditioning removes axis-aligned ill-conditioning but cannot do so when the ill-conditioning is cross-coupled or the search trajectory encounters saddles (i.e., landscapes with substantial negative Hessian spectral mass). We have not yet characterized which of these features each PINN sub-network falls into. This analysis is left for future work. 

\begin{table}[!ht]
\centering
\caption{Analytical Pretraining Loss Comparison Across sub-neural-networks}
\label{tab-pretrainingResults}
\begin{tabular}{lrrrc}
\hline
\textbf{Sub-network} & \textbf{Adam} & \textbf{DSO-EMA} & \textbf{DSO V2 Hybrid} & \textbf{Improvement Factor} \\
\hline
\emph{$\varphi_{net}$} & $3.09 \times 10^{-2}$ & $1.10 \times 10^{-1}$ & $\mathbf{4.68 \times 10^{-3}}$ & $23 \times$ \\
\emph{$cv_{net}$} & $\sim 10^{-10}$ & $1.87 \times 10^{-10}$ & $\mathbf{8.84 \times 10^{-13}}$ & $211 \times$ \\
\emph{$carrier_{net}$} & $\sim 10^{-8}$ & $7.10 \times 10^{-8}$ & $\mathbf{2.26 \times 10^{-9}}$ & $31 \times$ \\
\emph{$current_{net}$} & $\sim 10^{-10}$ & $1.74 \times 10^{-10}$ & $\mathbf{3.19 \times 10^{-13}}$ & $545 \times$ \\
\hline
\end{tabular}
\end{table}

\subsection{Empirical Activation Benchmark}

Activation function candidates for the loss residual training are compared in Table~\ref{tab-activationComparison}. The GELU provides superior Hessian stability and residual minimization over alternative activation function choices. One reason is that its second derivative is non-zero, unlike the ReLU activation function's. \begin{table}[!ht]
\centering
\caption{Quantitative Comparison of Activation Functions}
\label{tab-activationComparison}
\begin{tabular}{lcccc}
\hline
\textbf{Metric / Feature} & \textbf{ReLU} & \textbf{SIREN} & \textbf{Fourier+GELU} & \textbf{GELU (Selected)} \\
\hline
$C^\infty$ Smoothness & No & Yes & Yes & Yes \\
$\partial^2/\partial x^2 \neq 0$ & No & Yes & Yes & Yes \\
Monotone Physics Match & --- & No & Partial & Yes \\
Hessian Conditioning & --- & Poor & Good & Best \\
Input Dimension & 4 & 4 & 52 & 4 \\
Best Loss & --- & 183.85 & 68.1 & \textbf{6.35} \\
\hline
\end{tabular}
\end{table}

\subsection{Physical Validation Diagnostics and Experimental \texorpdfstring{$I\text{--}V$}{I-V} Fitting}

\begin{table}[!ht]
\centering
\caption{Validation Metrics and Physical Criteria}
\label{tab-validationMetrics}
\begin{tabular}{lll}
\hline
\textbf{Metric} & \textbf{Formula / Definition} & \textbf{Pass Criterion} \\
\hline
Curve Fit $R^2$ & $1 - \frac{\sum(I_{\mathrm{pred}} - I_{\mathrm{data}})^2}{\sum(I_{\mathrm{data}} - \bar{I})^2}$ & $> 0.95$ (Trace \& Retrace) \\
Interface Hysteresis & $\Delta_{\mathrm{interface}} = \langle C_v^{\mathrm{trace}}(0,t) - C_v^{\mathrm{retrace}}(0,t) \rangle_V$ & $> 0$ \\
Branch Ordering & Voltage rise thresholds & $V_{\mathrm{rise,trace}} < V_{\mathrm{rise,retrace}}$ \\
Depletion Width Ratio & $W_{\mathrm{learned}} / W_{\mathrm{expected}}$ where $W_{\mathrm{exp}} = \sqrt{\frac{2\varepsilon_{\mathrm{STO}} V_{bi}}{qN_D}}$ & $0.1 < \text{ratio} < 10$ \\
Charge Density Ratio & $|\rho_{\mathrm{learned}}| / (q N_D)$ where $\rho_{\mathrm{learned}} = -\varepsilon_{\mathrm{STO}} \frac{d^2\varphi}{dx^2}$ & $< 100$ \\
\hline
\end{tabular}
\end{table}

\begin{wrapfigure}{r}{0.42\textwidth}
\centering
\vspace*{-2em}
\includegraphics[width=0.4\columnwidth]{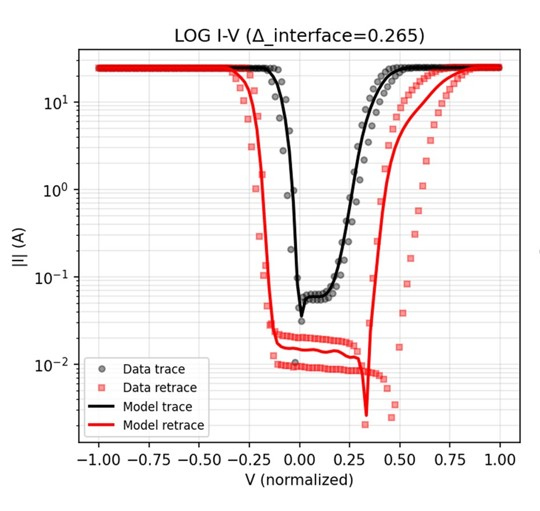} \vspace*{-2em}
\caption{Log-scale $I\text{--}V$ validation against experimental data (S04N dataset).}
\label{fig-iv-validation}
\vspace*{-2em}
\end{wrapfigure}

Table~\ref{tab-validationMetrics} summarizes the experiments confirming that predicted fields adhere strictly to physical semiconductor transport laws, and outputs are evaluated against pre-defined physical criteria. Oxygen partial pressure $P(\mathrm{O}_2)$ during growth directly modulates oxygen vacancy concentration in $\mathrm{SrTiO}_3$, inducing a potential drop $\Delta V = \sigma t / \varepsilon_{\mathrm{STO}}$ and corresponding Schottky barrier shifts. The PINN maps growth parameters to continuous internal fields and extracts macro-level device observables ($V_{\mathrm{up}}$, $V_{\mathrm{dn}}$). Experimental $I\text{--}V$ curves were acquired via conductive atomic force microscopy (c-AFM) \cite{kollias2025oxygen}, where trace and retrace sweeps correspond to increasing and decreasing voltage ramps, respectively. Figure~\ref{fig-iv-validation} illustrates how the framework achieves tight agreement across Schottky turn-on, hysteresis separation, and reverse-bias saturation regimes. The quantitative diagnostic results for the benchmark dataset are detailed in Table~\ref{tab-v2-metrics}. High $R^2$ coefficients ($>0.96$) validate the accuracy of inverse parameter extraction. A positive interface hysteresis metric ($\Delta_{\mathrm{interface}} = 0.265$) confirms vacancy-driven barrier modulation, while physical ratios ($W_{\mathrm{ratio}}=1.52$ and $\rho_{\mathrm{ratio}}=46.5$) confirm Poisson consistency and smooth depletion dynamics. 
\begin{table}[!ht]
\centering
\caption{All Validation Tests Passed Using S04N Dataset.}
\label{tab-v2-metrics}
\resizebox{\textwidth}{!}{%
\begin{tabular}{lccccrrrrr}
\hline
\textbf{Metric} & $R^2$ (trace) & $R^2$ (retrace)  & $\Delta_{\mathrm{interface}}$ & Trace rises first & Correct $C_v$ direction & Dynamic range  & $\Delta E_c$ (band offset) & $W_{\mathrm{ratio}}$  & $\rho_{\mathrm{ratio}}$\\
\textbf{Value}   & 0.9923& 0.9607& 0.265 & Yes & Yes &  $\sim 3$ decades & 0.35\,eV & 1.52 & 46.5\\
\textbf{Threshold} & $> 0.95$ & $> 0.95$ & $> 0$  & Req & Req & $\ge 2$ decades & 0.35\,eV  & $0.1$--$10$ & $< 100$ \\
\end{tabular}
}
\end{table}

\subsection{Discussion: PINN vs. COMSOL Solution Quality}
\label{sec-pinn-vs-comsol}

 Traditional finite-element solvers (e.g., COMSOL) utilize Newton--Raphson optimizer over sparse Jacobians $\mathbf{J} \in \mathbb{R}^{n \times n}$. Solver failures frequently occur due to exponential carrier concentration dependencies, severe spatial multiscale stiffness, and highly non-linear coupling between mobile vacancy concentrations and potential profiles \cite{kollias2018thesis}. The computational complexity per $I\text{--}V$ curve for solvers that use the Newton based optimizer is on the order of $\mathcal{O}(n^3)$. By contrast, PINN forward inference (as any neural network) is linear in the number of network parameters $N_\theta$ ($\mathcal{O}(N_\theta)$). Table~\ref{tab-pinn-vs-comsol} details solution quality metrics. From a computational physics perspective, the PINN surrogate resolves several fundamental bottlenecks inherent to conventional finite-element modeling tools like COMSOL \cite{kollias2025oxygen,kollias2018thesis}. Rather than approximating resistive switching hysteresis through unphysical static discrete zone shifts \cite{kollias2025oxygen,kollias2018thesis}, the neural surrogate models continuous ionic redistribution directly via dynamic latent variables ($\delta_{C_v}$). While standard grid-based discretizations require manual operator splitting, explicit Jacobian construction, and aggressive spatial re-meshing to resolve extreme multiscale Debye screening layers \cite{kollias2025oxygen,kollias2018thesis}, our architecture unifies coupled ionic, carrier, and electrostatic potential fields, represented as continuous functions $\varphi(x, t, V) = \varphi_{net}(x, t, V, C_v)$, into a single self-consistent computational graph evaluated pointwise via automatic differentiation. 

\begin{table}[!ht]
\centering
\caption{Comprehensive Solution Quality Comparison: PINN vs. COMSOL}
\label{tab-pinn-vs-comsol}
\begin{tabular}{lll}
\hline
\textbf{Aspect} & \textbf{COMSOL Multiphysics} & \textbf{PINN Framework} \\
\hline
$I\text{--}V$ Fit & Qualitative only & Quantitative ($R^2 = 0.992$ trace, $0.961$ retrace) \\
Hysteresis Representation & Discrete 5\,nm fixed zones & Continuous $C_v(x,t,V,b)$; $\Delta_{\mathrm{intf}} = 0.265$ \\
Band Offset $\Delta E_c$ & Fixed static input & Validated output parameter (0.35\,eV) \\
Depletion Width Extraction & Not reported & Explicitly extracted ($W_{\mathrm{ratio}} = 1.52$) \\
Poisson Consistency & Resolved at mesh nodes only & Continuous pointwise evaluation ($\rho_{\mathrm{ratio}} = 46.5$) \\
Spatial Domain & Artificially truncated ($10\,\mu\text{m}$) & Full physical scale ($380\,\mu\text{m}$) \\
Validation Status & Incomplete / Non-convergent & All physical metrics passed \\
\hline
\end{tabular}
\end{table}

The forward pass is fully differentiable, key physical transport and interface parameters ($R_{\mathrm{series}}$, $n$, $\beta$, $\delta_{C_v}$) are optimized end-to-end alongside network weights via backpropagation, eliminating the computationally prohibitive outer-loop parameter sweeps required by forward-only numerical solvers \cite{kollias2018thesis}. Thus, the high mesh-generation overhead, spatial-domain truncation, and solver non-convergence risks, associated with Newton--Raphson iterations on ill-conditioned problems in COMSOL \cite{kollias2025oxygen,kollias2018thesis}, are replaced by rapid, scalable GPU-based forward inference.

\section{Conclusion and Future Work}
\label{sec-conclusion}

Traditional COMSOL models of $\mathrm{Pt/STO/Si}$ heterostructures require artificial spatial truncation and external parameter sweeps, which frequently suffer from solver convergence issues \cite{kollias2018thesis}. The proposed PINN surrogate addresses these limitations by providing a complete, self-consistent PDE solution. The model delivers strong quantitative agreement with experimental data ($R^2 > 0.96$), continuous representation of internal output fields, non-truncated spatial domains ($380\,\mu\text{m}$), complete compliance across all physical validation metrics, and substantially faster inference than traditional finite-element solvers, whose per-curve cost scales as $\mathcal{O}(n^3)$ against the PINN's $\mathcal{O}(N_\theta)$ forward pass (Section~\ref{sec-pinn-vs-comsol}). Empirical NTK and Hessian--Gauss--Newton analyses \cite{podorozh2026sirenntk} may improve cubic regularized, recurrence based hybrid optimizers, such as the ARC-block family \cite{podorozh2026blockarcphi1}, by characterizing near-zero spectral bulk and informative Gauss--Newton outliers, distinguishing Gauss--Newton dominated from residual curvature dominated Hessian structure, and using $H-J^\top J$ ($H$ - Hessian, $J$ - Jacobian) as a candidate online signal for switching optimization modes. Optimizers whose step rule is chosen based on the analysis of this kind may further increase the fidelity of approximation of a PDE solution function. Building upon these results, future extensions will focus on extending the cascaded PINN design to 2D and 3D memristive crossbar arrays, incorporating thermal drift and self-heating multi-physics dynamics, and integrating the differentiable surrogate directly into closed-loop neuromorphic hardware design and real-time parameter extraction pipelines.

\bibliographystyle{unsrt}
\bibliography{NTheo,Tesic,RP31,PINN}

\end{document}